# Wherefore Quantum Mechanics?[†]

## Stephen Boughn[*]

Department of Physics, Princeton University, Princeton NJ
Departments of Physics and Astronomy, Haverford College, Haverford PA

## Abstract

After the development of a self-consistent quantum formalism nearly a century ago, there ensued a quest to understand the often counterintuitive predictions of the theory. These endeavors invariably begin with the assumption of the *truth* of the mathematical formalism of quantum mechanics and then proceed to investigate the theory's implications for the physical world. One of the outcomes has been endless discussions of the quantum measurement problem, wave/particle duality, the non-locality of entangled quantum states, Schrödinger's cat, and other philosophical conundrums. In this essay, I take the point of view that quantum mechanics is a mathematical model, a human invention, and rather than pondering what the theory implies about our world, I consider the transposed question: what is it about our world that leads us to a quantum mechanical model of it? One consequence is the realization that discrete quanta, the quantum of action in particular, leads to the wave nature and statistical behavior of matter rather than the other way around.

## Preface

Richard Feynman famously declared, "I think I can safely say that nobody really understands quantum mechanics."[1] Sean Carroll decried the persistence of this sentiment in a recent opinion piece entitled "Even Physicists Don't Understand Quantum Mechanics: Worse, they don't seem to want to understand it."[2] No one doubts the efficacy of quantum theory. The "understanding" to which these physicists refer is an acceptable ontology of the theoretical constructs of quantum mechanics. A typical query might be, "Do wave functions constitute a true description of physical reality?"[3]  For a pragmatist like me, this smacks more of metaphysics than physics.  I propose a more empirical question: "What is it about our world that has led us to a quantum mechanical description of it?"  The purpose of my essay is to answer this question or at least shed some light on it.  Before I tell you what I mean by the question, let me tell you what I don't mean. I am not referring to how experimental observations of the details of atomic spectra, the spectrum of blackbody radiation, the photoelectric effect, the heat capacity of solids, etc, led to the Rutherford-Bohr atomic model and eventually to Heisenberg's matrix mechanics and Schrödinger's wave mechanics.  The only immediate answer this would seem to provide to the above question is something like "Quantum mechanics simply describes the observed behavior of Nature", a rather circular response that isn't particularly helpful to me in my quest to understand why this is the case.  I am definitely not referring to the axiomatic approach of Dirac and von Neumann in their development of the mathematical foundations of quantum mechanics and the accompanying interpretations of the formal constructs of the theory.  Axiomatic approaches in which the equations of quantum mechanics are derived from something more fundamental, for example symmetry principles, similarly don't provide the sort of answer I seek.  These two comprise the standard textbook presentation of quantum mechanics and have been enormously useful to us in our quest to understand nature not to mention in facilitating extraordinary advances in technology.  But I want something more.  On the other hand, perhaps we should simply accept Feynman's appraisal and not seek a deeper understanding but rather follow the admonition, "Shut up and calculate!"[4]

---

[*] sboughn@haverford.edu

[1] The 1964 Messenger Lectures at Cornell University, Lecture #6.

[2] NY Times, Sunday Review, Sept. 7, 2019

Even so, many have pursued precisely this quest: de Broglie and Bohm's pilot wave theory; Hugh Everett's many world interpretation; John Bell and his argument for quantum non-locality; the spontaneous wave function collapse theory of Ghirardi, Rimini, and Weber; Griffiths, Gell-Mann, and Hartle's consistent histories interpretation; quantum Bayesianism; and many others.  An underlying assumption of all these efforts is that the world is governed by a set of comprehensible universal laws.  Once these laws are discovered, it seems eminently reasonable to pursue an understanding of them.  Because fundamental laws inevitably take the form of formal mathematical models, it's not surprising that investigations of their meanings involve highly theoretical analyses as evidenced by the above examples. Such efforts don't

---

[3] This was the question posed by Einstein in the famous EPR paper. (Einstein *et al.* 1935)

[4] The full David Mermin quote is "If I were forced to sum up in one sentence what the Copenhagen interpretation says to me, it would be 'Shut up and calculate!'" (Mermin 1989)

resonate with me and don't constitute the sort of understanding that I have in mind. Perhaps this is because I'm an experimental physicist with a disposition for pragmatism. My philosophy of physics[5] is much more empirical than theoretical. For me physical theories are not laws of nature but rather human creations, models that we invent to help us make sense of our experiences of the world.

So just what do I mean by the question "What is it about our world that has led us to a quantum mechanical description of it?" Well, I'm not quite sure. Maybe I'm searching for something analogous to how the constancy of the speed of light leads us to the theory of special relativity or to Heisenberg's attempt at a fundamental understanding of quantum mechanics that led to his seminal paper on the uncertainly principle. (I'll return to Heisenberg's argument in Sections 1 and 4.) An ancient example is how pre-Socratic philosophers were lead to atomism. (See Section 1.) Because I'm endeavoring to discover the empirical basis for our quantum theoretical model, it's incumbent upon me to avoid, as much as possible, reference to theoretical constructs. This is easier said than done and you will find that I haven't been entirely successful in this regard. Planck's quantum of action, $h$, will figure prominently in the following discussion and it's difficult to discuss it without relying on a mathematical model, which runs counter to my intent of avoiding reference to formal theories. On the other hand, I'm not particularly embarrassed by the vagueness that this necessarily introduces into some of my arguments.[6] As Heisenberg noted about Bohr's Copenhagen interpretation (Stapp 1972),

> Besides that it may be a point in the Copenhagen interpretation that its language has a certain degree of vagueness, and I doubt whether it can become clearer by trying to avoid this vagueness.

My approach to the question I've posed will be more heuristic than formal and theoretical. In addition to providing a more empirical understanding of quantum mechanics, I hope that such an approach will make some of the mysteries of the theory more palatable and, perhaps, will help to dispel some of the intractable quantum conundrums, like the measurement problem. If you do not find this approach of interest, I certainly won't be offended if you don't continue reading my essay. ☺

---

[5] I've expressed this view in more detail in "On the Philosophical Foundations of Physics: An Experimentalist's Perspective" (Boughn 2019).
[6] In fact, the relation of experiments to theory is fundamentally vague even in the context of classical physics as will be discussed in Section 5.

# 1 Introduction

To be sure, probing well-established theories to discover previously unknown physical phenomena is a tried-and-true enterprise in physics. Examples include: electromagnetic radiation; neutron stars; antimatter; Bose-Einstein condensation; neutrinos; gravitational waves; the Higgs boson; and many more. In no way is my intention to minimize these wonderful predictions and their subsequent experimental confirmation. They all involve specific observable effects predicted by physical theories. Such predictions are possible because of the well-defined meaning of statements made within the context of the theories. However, the quantum conundrums alluded to in the preface are paradoxes of a different sort. They include wave function collapse, action at a distance, Schrödinger's cat, the quantum/classical divide, and perhaps even wave/particle duality. These all arise from trying to conjure new meanings of theoretical constructs, meanings that were not included in the foundations of the theory. I've discussed the futility of such efforts elsewhere (Boughn & Reginatto 2013, Boughn 2019). One purpose of the present essay is to redirect those queries from questions about the metaphysics of the quantum formalism to the question, what is it about the physical world that has led us to invent that formalism.

Heisenberg was, perhaps, the first to address such a question. His seminal 1927 paper employed a gedanken experiment, in the form of a gamma ray microscope, to establish an *indeterminacy relation*, $\delta x \delta p \sim h$. In the abstract he remarked, "This indeterminacy is the real basis for the occurrence of statistical relations in quantum mechanics." The title of the paper, "Über den anschaulichen Inhalt der quantentheoretischen Kinematik und Mechanik", was translated by Wheeler & Zurek(1983) as "The Physical Content of Quantum Kinematics and Mechanics". The German phrase *anschaulichen inhalt* has been variously translated as *physical content* (Wheeler & Zurek 1983), *intuitive contents* (Miller 1994), and *perceptual content* (Cassidy 1992). It is clear that Heisenberg was trying to render quantum mechanics more *anschaulich* by appealing to an empirical property of nature, in this case demonstrated with his gamma ray microscope.[7]

Before proceeding, it's necessary to define what is meant by "a quantum mechanical description" of nature. The formalism of quantum mechanics is based on specifying the quantum state of a system in terms of either the Schrödinger wave function or Hilbert state vector and a Hamiltonian that describes the evolution of the quantum state. The meanings of these theoretical constructs derive from the way physicists use them. Perhaps the simplest way of expressing this meaning is associated with the Copenhagen interpretation of Bohr and Heisenberg. Consider a quantum system characterized by a wave function $\psi(x)$ and suppose that a property characterized by an operator $\boldsymbol{A}$ is to be measured for this system. Further suppose that $\phi_k(x)$ represent the eigenfunctions of $\boldsymbol{A}$ such that $\boldsymbol{A}\phi_k = a_k \phi_k$ where $a_k$ are the eigenvalues of $\boldsymbol{A}$ and constitute all possible results for measurements of the given property. According to the Born rule, the probability that a measurement characterized by $\boldsymbol{A}$ will yield the value $a_k$ is given by $|\int dx\, \psi\, \phi_k|^2$. These relations encompass what we mean by the quantum state and quantum operator. The statistical aspects of quantum mechanics follows directly from the Born rule while the wave/particle duality implied by the Copenhagen interpretation is a consequence of this rule and the Hamiltonian operator that governs the structure and evolution of the wave function. Of course, quantum mechanics proper encompasses far more than this simple example; however, the meanings of all the theoretical constructs (like $\psi(x)$ and $\boldsymbol{A}$) are derived from similar considerations. The question remains, why are we lead to such a description of the natural world.

---

[7] We'll come back to Heisenberg's paper in Section 4.

First, let's list a few of the counterintuitive aspects of quantum mechanics. Perhaps the most egregious example is the observation that electrons and light exhibit both the properties of particles and the properties of waves, behavior that is anathema to classical physics. The Schrödinger wave function certainly pointed to the wave properties of particles and led De Broglie and Schrödinger to interpret quantum waves as continuous distributions of matter. It was Einstein and Born who introduced statistics to quantum mechanics by suggesting that the wave function provides a statistical measure of where a particle of radiation or matter might be located [8] (Stone 2013). After 1930, the Copenhagen interpretation of Bohr and Heisenberg was generally accepted; although, Bohr and Heisenberg often emphasized different aspects of the interpretation and there has never been complete agreement as to its meaning even among its proponents (Stapp 1972). The Copenhagen interpretation dealt with the incongruous dual wave and particle properties by embracing Bohr's principle of complementarity according to which complementary features of physical systems can only be accessed by experiments designed to observe one or the other but not both of these features. For example, one can observe either the particle behavior or wave behavior of electrons but not both at the same time. In addition, the waves implicit in Schrödinger's equation were interpreted in terms of probability amplitudes for the outcomes of experiments. Finally, in order to facilitate the relation of quantum formalism to experimental results, the Copenhagen interpretation emphasized that the description of experiments, which invariably involve macroscopic apparatus, must be expressed in classical terms. These aspects of quantum theory are familiar to all beginning students of quantum mechanics; however, many students harbor the uneasy feeling that something is missing. How can an electron in some circumstances exhibit the properties of a particle and at other times exhibit the properties of a wave? How is it that the primary theoretical constructs of quantum mechanics, the Schrödinger wave functions or Hilbert state vectors, only indicate the probability of events? Quantum mechanics itself does not seem to indicate that any event actually happens. Why is it that experiments are only to be described classically? Where is the quantum/classical divide between the quantum system and the classical measurement and what governs interactions across this divide? In fact, these sorts of questions are raised not only by neophyte students of quantum mechanics but also by seasoned practitioners. In actuality, the question of how to interpret quantum theory has never been fully answered and new points of view are still being proffered.

---

[8] One can appreciate the irony since it was the statistical aspects of quantum mechanics that most bothered Einstein.

Seeking answers to these questions is a long and venerated enterprise that has been pursued by philosophers and theoretical physicists alike. These pursuits, far from being idle avocations, have resulted in important contributions to the foundations of quantum mechanics. However, the purpose of this essay is to address a different, more epistemological question, "What is it about the physical world that leads us to a quantum theoretic model of it?", a question that is still pondered by some physicists and philosophers and certainly by many physics students when they first encounter quantum mechanics. Most of the latter group eventually come to some understanding, perhaps via the ubiquitous Copenhagen interpretation, and then proceed according to the "shut up and calculate" maxim. One modest aim of this essay is to provide such students with a heuristic perspective on quantum mechanics that might enable them to proceed to calculations without first having to "shut up".

## 2 What's Quantized?

Let us begin by asking where the 'quantum' in quantum mechanics comes from. What is it that's quantized? That matter is composed of discrete quanta, atoms, was contemplated by Greek philosophers in the $5^{th}$ century B.C. (Berryman 2011) and the idea continued to be espoused through the $18^{th}$ century. Even though it wasn't until the $19^{th}$ and early $20^{th}$ centuries that the existence of atoms was placed on a firm empirical basis, it's not difficult to imagine what led early philosophers to an atomistic model. Perhaps the primary motivation, an argument that still resonates today, was to address the puzzle of change, the transformation of matter. This was often expressed by the assertion that things cannot come from nothing nor can they ever return to nothing. Rather, creation, destruction, and change are most simply explained by the rearrangement of the atomic constituents of matter. In his epic poem, *De rerum natura* (On the Nature of Things, circa 55 BC), Lucretius[9] explained (translation by R. Melville 1997),

> ...no single thing returns to nothing but at its dissolution everything returns to matter's primal particles...they must for sure consist of changeless matter. For if the primal atoms could suffer change...then no more would certainty exist of what can be and what cannot...Nor could so oft the race of men repeat the nature, manners, habits of their parents.[10]

While it took nearly 2500 years, this conjecture of the atomists was largely justified.

One might also reasonably ask, "Are there other aspects of nature that are quantized?" It's no coincidence that during the same period that saw the confirmation of the atomic hypothesis, there appeared evidence for the discrete nature of atomic interactions. Perhaps the first clues were the early $19^{th}$ century observations by Wollaston and Fraunhofer of discrete absorption lines in the spectrum of the sun and in 1859 the subsequent identification of emission lines in the spectra of elements in the laboratory by Kirchoff and Bunsen. In 1888, Rydberg was able to relate the wavelengths of these discrete spectral lines to ratios of integers. Boltzmann introduced discrete energy as early as 1868 but only as a computational device in statistical mechanics. It was in 1900 that Planck found he must take such quantization more seriously in his derivation of the Planck black body formula (Badino 2009). A decade later Jeans, Poincaré, and

---

[9] Lucretius was a disciple of the Greek atomist Epicurus and his predecessors Democritus and Leucippus (Berryman 2011).

[10] One might balk at attributing human characteristics to atoms; however, microscopic genes composed of atoms certainly also qualify as (ordinarily) changeless primal particles.

Ehrenfest demonstrated that the discreteness of energy states, which source black body radiation, follows from the general morphology of the spectrum and is not the consequence of precisely fitting the observed spectral data (Norton 1993). In 1905 Einstein introduced the notion of quanta of light with energies $E$ that depended on frequency $\nu$ with precisely the same relation as introduced by Planck[11], $E = h\nu$, where $h$ is Planck's constant. He then used this relation to explain qualitative observations of the photoelectric effect.[12] In 1907 it was again Einstein who demonstrated that energy quantization of harmonic oscillators explained why the heat capacities of solids decrease at low temperatures. Finally, Bohr's 1913 model of discrete energy levels of electrons in atoms explained the spectral lines of Kirchoff and Bunsen as well as resolved the conflict of Maxwell's electrodynamics with the stability of Rutherford's 1911 nuclear atomic model.

In a 1922 conversation with Heisenberg (Heisenberg 1972), Bohr expressed an argument for the discreteness of atomic interactions that harkened back to the ancient Greeks' arguments for atoms (and to the Lucretius quote above). Bohr based his argument on the stability of matter, but not in the sense just mentioned. Bohr explained,

> By 'stability' I mean that the same substances always have the same properties, that the same crystals recur, the same chemical compounds, etc. In other words, even after a host of changes due to external influences, an iron atom will always remain an iron atom, with exactly the same properties as before. This cannot be explained by the principles of classical mechanics...according to which all effects have precisely determined causes, and according to which the present state of a phenomenon or process is fully determined by the one that immediately preceded it.

In other words, in a world composed of Rutherford atoms, quantum discreteness is necessary in order to preserve the simplicity and regularity of nature. Bohr's 'stability' and Lucretius's 'repeatability' clearly refer to the same aspect of nature.

It might appear from the examples given above that energy is the key dynamical quantity that must always come in discrete quanta. However, there are problems with this demand. For one thing, there is no fundamental constant in physics, such as the speed of light c, Planck's constant h, or Newton's constant G, that has the units of energy.[13] In addition, it's straightforward to demonstrate that energy is not quantized in all situations. Even in the context of quantum mechanics, a free particle can assume any of a continuum of values of energy and momentum and there are many interactions (e.g., Rutherford and Compton scattering) for which a particle's energy and momentum change by arbitrarily small increments. Certainly there are many instances of the quantization of energy $E$, momentum $p$, and even position $x$; however, the values of these quanta depend on the specifics of the system and have a wide range of values. For example, the energy and momentum in a monochromatic beam of photons are quantized in units of $h\nu$ and $h\nu/c$ but the values of these quanta depend on the frequency of the photons and are unconstrained; they can take on any value between 0 and $\infty$, which again argues against their primal status.

---

[11] It is interesting that in 1899, the year before his seminal Planck's Law paper, Planck introduced the constant that bears his name (although he gave it the symbol $b$, not $h$) from Paschen's fit of spectral data to Wien's Law. Even then, he identified it as a fundamental constant of nature along side $e$, $c$, and $G$. (Planck 1899)

[12] Einstein's quantitative prediction was confirmed by the 1914 experiments of Robert Millikan.

Angular momentum would seem to be a more likely candidate for the quantum that characterizes atomic interactions. Indeed, there is a fundamental physical constant that has the units of angular momentum, namely Planck's constant, $h$. In fact, a key tenant of Bohr's atomic model was the quantization of angular momentum. On the other hand, that such a specific quantity as angular momentum should occupy this primal status might give one pause. To be sure, in the context of standard quantum mechanics, the quantum state of a particle can always be expressed as a linear combination of angular momentum eigenfunctions (spherical harmonics), with eigenvalues $\boldsymbol{L}_z = m\hbar$ and $\boldsymbol{L}^2 = l(l+1)\hbar^2$, where $\boldsymbol{L}$ and $\boldsymbol{L}_z$ are the angular momentum operator and its $z$ component, $m$ and $l$ are integers, and $\hbar \equiv h/2\pi$. However, recall that the *modus operandi* of this essay is to avoid drawing inferences from quantum formalism but instead to identify empirical properties of nature that lead us to a quantum mechanical description of it.

An alternate approach might be to consider the quantization of some specific combination of $E$, $p$, and $x$. In fact, Heisenberg's indeterminacy relation (Heisenberg 1927), $\delta x \delta p \sim \hbar$, points to the product of position and momentum as such a combination and we will see that this is, indeed, prescient. It is not inappropriate to invoke indeterminacy in this context because Heisenberg arrived at his relation, not from quantum mechanics, but rather from an empirical

---

[13] To be sure, the Planck energy, $\sqrt{\hbar c^5/G}$, is fundamental but is much too large ($2 \times 10^9$ Joules) to be relevant on atomic scales.

gedanken experiment involving a gamma ray microscope, as will be discussed later.

In 1912, Nicholson proposed that the angular momentum of an electron in orbit about an atomic nucleus is quantized and the following year Ehrenfest argued that the unit of this quantum is $\hbar \equiv h/2\pi$. In the same year, Bohr incorporated these ideas into his model of an atom, a model that provided a successful explanation of the spectrum of atomic hydrogen. Even though $\hbar$ has the same units as angular momentum, recall that I expressed skepticism that angular momentum characterizes the fundamental quantum of interaction. Wilson, Ishiwara, Epstein, and Sommerfeld soon replaced Nicholoson/Ehrenfest/Bohr quantization with the notion that Hamilton-Jacobi action variables $J_k$ (for periodic systems), which also have the same units as $h$, are quantized (Whittaker 1989). That is, $J_k \equiv \oint p_k dx_k = nh$ where $p_k$ is the momentum conjugate to the coordinate $x_k$, $n$ is an integer, and the integral is taken over one cycle of periodic motion. On the other hand, the quantization of action depends on the coordinates employed, which again seems unacceptable for a fundamental principle.

In 1917 Einstein gave a new interpretation of the quantization conditions by demonstrating that they followed from the requirement that Hamilton's principle function *S* is multivalued such that the change in *S* around any closed curve in configuration space is an integer times Planck's constant, i.e., $\oint p_k dx_k = \oint (\partial S/\partial q_k) dq_k = \oint dS = nh$ (Stone 2005). This geometric expression no longer depends on the choice of coordinates. It is straightforward to demonstrate that the quantization of energy for harmonic oscillators and photons, as well as angular momentum quantization in the Bohr atom, follow from the quantization of the Hamilton-Jacobi action, $\oint dS = nh$.

Even though the Hamiltonian action integral is a quantity defined in the context of classical mechanics, it is certainly a theoretical construct (as are energy, momentum, and angular momentum for that matter) and therefore runs counter to my desire to construct an empirical basis for our quantum theoretical model, a shortcoming of which you've been forewarned. However, the description of nearly any observation necessarily requires some sort of theoretical basis and a classical description is perhaps the lesser of evils. Also, the above argument is semi-classical and, as such, it's difficult to imagine how it could provide a firm foundation for modern quantum theory. However, the reader should be reminded that the purpose of this essay is not an axiomatic derivation of quantum mechanics from fundamental principles but rather to acquire insight into the quantum world and thus address the question, "What is it about the physical world that led us to a quantum theoretic model of it?" I now continue with this task.

## 3 Quantization and Waves

Duane (1923), Breit (1923) and Compton (1923) applied the quantization of action to the interaction of x-ray photons with an infinite, periodic crystal lattice and were able to obtain Bragg's law of diffraction without directly invoking the wave nature of x-rays. A somewhat simpler case is that of photons incident upon an infinite diffraction grating. Figure 1 is a replica of the schematic diagram in Breit's 1923 paper where $h\nu/c = p_\gamma$ is the momentum of the incident photon, $G$ is the diffraction grating, $D_0, D_{\pm 1}, D_{\pm 2}, ...$ are the positions of the slits of the grating, $\theta$ is the scattering angle, and $P$ is the transverse momentum of the emergent photon. Now assume that the momentum transferred from the radiation to the grating is governed by the quantization condition $\oint p\,dx = nh$ where $p$ is in a direction parallel to its surface and the integral is taken over the transverse distance necessary to bring the system back to its original condition, i.e., the line spacing $d = D_{k+1} - D_k$. In this case, the average momentum transferred to the grating is $\langle p \rangle = \oint p\,dx / \oint dx = nh/d$ and by conservation of momentum this must also be the magnitude of the transverse momentum transferred to an incident photon, i.e., $P = \langle p \rangle$. If photons are incident perpendicular to the plane of the grating, then the allowed angles at which they are transmitted through the grating are given by $sin\theta_n = \langle p \rangle / p_\gamma.$[14] Thus, $sin\theta_n = nh/p_\gamma d$, which is the relation for diffraction (interference) of a wave with wavelength $\lambda = h/p_\gamma$. No specific reference to the wave nature of the photons is necessary. Breit (1923) and Epstein & Ehrenfest (1924) extended these results to finite width, single and multiple slit interference patterns. Thus, the quantization condition $\oint p dx = nh$ leads to the interference properties of photons without directly invoking their wave nature. It is curious that none of these authors extended their analyses to the case of electrons scattered from crystals, a process that should obey the same quantization condition. If they had, they might have predicted that $\lambda = h/p$ and the wave nature of electrons prior to de Broglie's 1924 thesis and Davisson & Germer's and Thomson's 1927 electron diffraction experiments. The analyses of Duane *et al.* provide seminal illustrations of a direct path from the quantization of action to the wave behavior of particles and photons. As such, they lend credence to the notion that there is a primal relation between the quantization of dynamical properties and the dual wave-particle behavior of quantum systems.

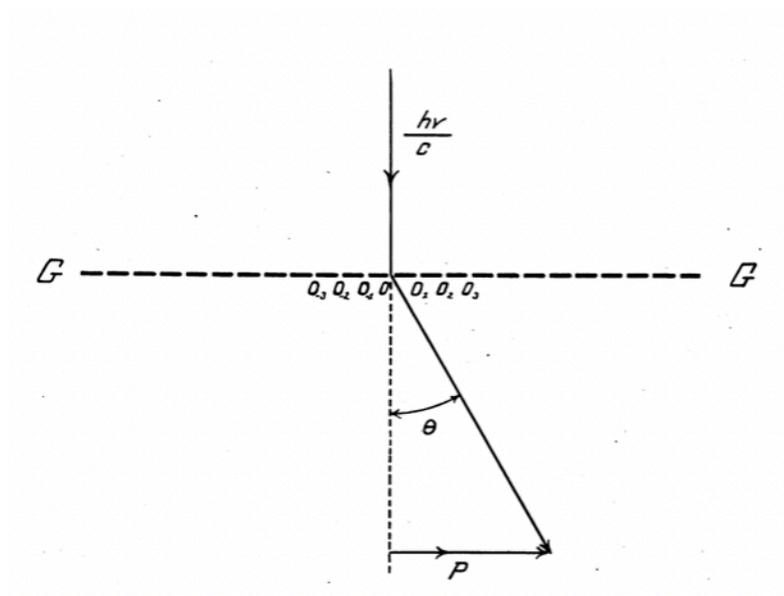

Figure 1: Photon scattered from an infinite diffraction grating from Breit (1923)

## 4 Physics and Probability

Another major conundrum of quantum mechanics is the fundamental role of probability in the theory.[15] The probabilities are taken to apply to the outcomes of possible observations of a system even though some of the observations are mutually exclusive (Bohr's principle of complementarity). This seems to fly in the face of our classical notion that physical systems should be completely describable in isolation, prior to and independent of any observation. How is it that the specification of mere probabilities can possibly constitute a fundamental description of a physical system and if so, how can such a description possibly provide a complete description of reality?[16] In 1927 Heisenberg proposed the indeterminacy relation, $\delta x \delta p \sim h$, that bears his name. It was his contention that "this indeterminacy is the real basis for the occurrence of statistical relations in quantum mechanics."(Heisenberg 1927) He arrived at the concept by considering a gedanken experiment in the form of a gamma ray microscope. Heisenberg reasoned that with such a microscope one could only determine an electron's position to within on the order of one gamma ray wavelength, $\delta x \sim \lambda$. But in doing so, one would impart to the electron an unknown momentum on the order of the momentum of the incident gamma ray, $\delta p \sim E_\gamma/c = h\nu/c = h/\lambda,$ and hence $\delta x \delta p \sim h.$[17] To the extent that the wave behavior of gamma rays follows from quantization, as demonstrated by Duane et al., the Heisenberg indeterminacy relation is a direct consequence of the quantum of action. Heisenberg demonstrated that this relation can also be determined directly from the formalism of quantum mechanics; however, our point here is that it is already evident from the quantization of action.

---

[14] Because the mass of the grating is very large, the momentum and energy of the scattered photon, $p_\gamma$ and $p_\gamma c$, do not change, a result that follows from the Compton effect discovered only a few months earlier.

[15] It was Einstein who first suggested that the intensity of electromagnetic waves was a measure of the probability of the location of photons. Born extended this notion to particles with a similar interpretation of the wave functions of Schrödinger's equation (Stone 2013).

[16] In fact, Einstein, Podolsky, and Rosen (1935) maintained "that the description of reality as given by a [quantum] wave function is not complete."

The Heisenberg uncertainty principle is one of the pillars of modern physics and his gamma ray microscope provides a particularly intuitive interpretation of the principle. However, there are other insightful gedanken experiments that are more directly tied to quantization. For example, suppose a particle is confined to be within a one-dimensional box (potential well) of width $l$ but is otherwise free, i.e., has constant momentum $p$ along the one dimension but in either direction. The motion of the particle will clearly be periodic with a spatial period $2l$ and the quantization condition is $\oint p dx = 2pl = nh.$ If the particle is in its ground state, $n = 1$ and $2pl = h.$ At any instant, the uncertainty in the particle's position is clearly $\delta x \sim l.$ The magnitude of the particle's momentum is known but it could be moving in either direction so the uncertainty in its momentum is $\delta p \sim h/l.$ Combining these two relations, we again arrive at Heisenberg's indeterminacy relation, $\delta x \delta p \sim h$. Of course, this particle is confined; however, if the box is opened, the particle is free to move in either direction. Immediately after the box is opened, the uncertainties in the position and momentum of the now free particle again satisfy the Heisenberg relation, $\delta x \delta p \sim h.$

The argument that Heisenberg gave to support his contention that his indeterminacy relation is the basis for the statistical relations in quantum mechanics is as follows (Heisenberg 1927),

> We have not assumed that quantum theory–in opposition to classical theory–
> is an essentially statistical theory in the sense that only statistical conclusions
> can be drawn from precise statistical data....Rather, in all cases in which
> relations exist in classical theory between quantities which are really exactly measurable, the corresponding exact relations also hold in quantum theory (laws of conservation of momentum and energy). But what is wrong in the sharp formulation of the law of causality, "When we know the present precisely, we can predict the future," is not the conclusion but the assumption. Even in principle we cannot know the present in all detail. For that reason everything observed is a selection from a plenitude of possibilities and a limitation on what is possible in the future.

---

[17] Heisenberg also argued that similar indeterminacy relations occurred for all conjugate pairs of observable quantities.

Another reason to concede to a statistical view of nature is the realization that this notion is not particularly foreign to classical physics. Certainly, statistical mechanics is one of the triumphal successes of classical physics. On the experimental side, careful consideration of uncertainties is always essential when comparing observations with theoretical predictions, either quantum or classical. In the classical case these uncertainties are usually viewed as experimental "noise" and left to the experimentalist to elucidate. However, this doesn't necessarily have to be the case. The Hamilton-Jacobi formalism provides an approach in which such uncertainties can be included in the fundamental equations of classical mechanics[18] (Hall and Reginatto 2005, 2016); although, it is usually far more convenient to deal with them in the analysis of a measurement rather than as fundamental facet of the theory. An interesting aside is that by combining the statistical Hamilton-Jacobi formalism of classical mechanics with the Heisenberg uncertainty relations, one can generate a plausible route to Schrödinger's equation and the concomitant wave nature of particles (Hall & Reginatto 2002, Boughn & Reginatto 2018). One can even construe statistical relations in classical physics in terms of classical indeterminacy relations $\delta x > 0$ and $\delta p > 0$ (Volovich 2011). In a very real sense, violations of these relations, namely $\delta x = 0$ or $\delta p = 0$, are just as inaccessible as a violation of the quantum mechanical uncertainty principle, $\delta x \delta p < \hbar/2$, an assertion to which any experimentalist will attest.[19] These arguments are certainly not intended to demonstrate that quantum mechanics and classical mechanics are compatible. Clearly, they are not. They are offered simply to emphasize that probability and statistics are fundamental to physics, both classical and quantum. Rather, the crucial difference between the two is the quantization of action that is primal in quantum physics but absent in classical physics. To be sure, it is for this reason that the statistical nature of quantum mechanics (via the Born rule) seems to be more fundamental than in classical mechanics.

---

[18] In fact, some experimental uncertainties are routinely included in quantum mechanical calculations expressed as the weightings in *mixed states*.

[19] Note that here I've replaced Heisenberg's $\delta x \delta p \sim \hbar$ with the usual $\delta x \delta p \geq \hbar/2$, which is derived from the corresponding quantum mechanical commutation relation.

Because the notion of discrete quanta of both matter and physical interactions figures so prominently in this essay, let me speculate about a connection between discrete quanta and a statistical model of nature. Heisenberg's argument certainly points in this direction; however, I want to be a bit more general. In everyday experience when we encounter discrete phenomena we often resort to probabilistic descriptions. Examples include flipping a coin, rolling dice, spinning a roulette wheel, and employing a Galton board to visualize a binomial (and approximately normal) statistical distribution. One might object that such descriptions are only statistical because we lack precise knowledge of the original state of the object but this objection already invokes Newtonian theory, which I'm endeavoring to avoid. I'm only drawing on "everyday experiences" for insight into how one might describe nature. Also, precise knowledge of the original state (in the Newtonian sense) is problematic because to achieve this one would have to interrogate the system via discrete interactions with another system whose state must also be precisely known, and so on *ad infinitum*, which brings us back to Heisenberg's argument based on his indeterminacy relation. Two 19th century microscopic examples are: Gregor Mendel's observations of the statistical behavior of the inheritance patterns of pea plants that led him to introduce the concept of discrete inherited units; and Boltzmann's introduction of discrete units of energy as a computational device in statistical mechanics. To me, these illustrations are an indication as to why it should be no surprise that a world characterized by discrete quanta might lead to a fundamentally statistical model of nature.

## 5 The Quantum/Classical Divide

The dual wave-particle nature of matter and radiation and the probabilistic character of the theory are not the only elements that exasperate beginning students of quantum mechanics. Another point of discomfort is the quantum/classical divide that the Copenhagen interpretation places between a quantum system and a classical measuring apparatus. Where is the divide and what physical interactions occur across the divide? This dilemma is predicated upon the supposition that experiments must be, or inevitably are, described by classical physics. Upon closer inspection, the assertion that classical physics adequately describes experiments is far from obvious. Bohr expressed the situation as follows (Bohr 1963):

> The decisive point is to recognize that the description of the
> experimental arrangement and the recordings of observations must
> be given in plain language, suitably refined by the usual terminology.
> This is a simple logical demand, since by the word 'experiment' we
> can only mean a procedure regarding which we are able to communicate
> to others what we have done and what we have learnt.

Stapp (1972) chose to emphasize this pragmatic view of classicality by using the word specifications, i.e.,

> Specifications are what architects and builders, and mechanics and
> machinists, use to communicate to one another conditions on the
> concrete social realities or actualities that bind their lives together.
> It is hard to think of a theoretical concept that could have a more
> objective meaning. Specifications are described in technical jargon
> that is an extension of everyday language. This language may
> incorporate concepts from classical physics. But this fact in no way
> implies that these concepts are valid beyond the realm in which they
> are used by technicians.

The bottom line is that descriptions of experiments are invariably given in terms of operational prescriptions or specifications that can be communicated to technicians, engineers, and the physics community at large. The formalism of quantum mechanics has absolutely nothing to say about experiments.

There have been many proposed theoretical resolutions to the problem of the quantum/classical divide but none of them seem adequate (e.g., Boughn & Reginatto 2013). One obvious approach is simply to treat the measuring apparatus as a quantum mechanical system. While perhaps impractical, no one doubts that quantum mechanics applies to the bulk properties of matter and so this path might, in principle, seem reasonable. However to the extent that it can be accomplished, the apparatus becomes part of the (probabilistic) quantum mechanical system for which yet another measuring apparatus is required to observe the combined system. Heisenberg expressed this in the extreme case, "One may treat the whole world as one mechanical system, but then only a mathematical problem remains while access to observation is closed off."(Schlosshauer & Camilleri 2008)

Ultimately, the dilemma of the quantum/classical divide or rather system/experiment divide is a faux problem. Precisely the same situation occurs in classical physics but apparently has not been considered problematic. Are the operational prescriptions of experiments part and parcel of classical theory? Are they couched in terms of point particles, rigid solid bodies, Newton's laws or Hamilton-Jacobi theory? Of course not. They are part of Bohr's "procedure regarding which we are able to communicate to others what we have done and what we have learnt." Therefore, it seems that the problem of the relation of theory and measurement doesn't arise with quantum mechanics but exists in classical mechanics as well. At a 1962 conference on the foundations of quantum mechanics, Wendell Furry explained (Furry 1962),

> So that in quantum theory we have something not really worse than we had in classical theory. In both theories you don't say what you do when you make a measurement, what the process is. But in quantum theory we have our attention focused on this situation. And we do become uncomfortable about it, because we have to talk about the effects of the measurement on the systems....I am asking for something that the formalism doesn't contain, finally when you describe a measurement. Now, classical theory doesn't contain any description of measurement. It doesn't contain anywhere near as much theory of measurement as we have here [in quantum mechanics]. There is a gap in the quantum mechanical theory of measurement. In classical theory there is practically no theory of measurement at all, as far as I know.

At that same conference Eugene Wigner put it like this (Wigner 1962),

> Now, how does the experimentalist know that this apparatus will measure for him the position? "Oh", you say, "he observed that apparatus. He looked at it." Well that means that he carried out a measurement on it. How did he know that the apparatus with which he carried out that measurement will tell him the properties of the apparatus? Fundamentally, this is again a chain which has no beginning. And at the end we have to say, "We learned that as children how to judge what is around us." And there is no way to do this scientifically. The fact that in quantum mechanics we try to analyze the measurement process only brought this home to us that much sharply.

Physicists have long since become comfortable with the relation between theory and measurement in classical physics so, perhaps, the quantum case shouldn't be regarded as particularly worrisome.

## 6 Back to Quanta

I began this essay with the question "What is it about the physical world around us that leads us to a quantum theoretic model of it?" and have tried to answer it by discussing the quantal character of the physical world along with the inevitability of the statistical nature of both quantum and classical physics. In addition, when compared with its classical counterpart, the relationship of theory and measurement in quantum mechanics doesn't seem all that unusual. I suspect that many of those who, like Feynman, lament not understanding quantum mechanics would claim that such notions as quantum interference, quantum non-locality, and wave-particle duality simply don't make sense. However, I hope I've convinced you that once one accepts the fundamental nature of the *quantum* these notions do make sense. On the other hand, the Greek atomists couldn't make sense of nature without quanta of matter and, as Bohr pointed out, this was also the case without discrete quanta of atomic interactions (see Section 2). In fact, in the absence of the notion of the quantum discreteness, it is the classical theory of nature, not the quantum theory, that makes no sense.

I hope these musings will provide some comfort to beginning students of quantum mechanics by providing a heuristic answer that bears upon the epistemological origin of wave-particle duality, the probabilistic interpretation of quantum formalism, and the somewhat elusive connection of theoretical formalism and measurements. Perhaps they will be afforded some solace as their credulity is strained by references to the quantum/classical divide, the collapse of the wave function, and the spooky action at a distance of entangled quantum systems. I personally suspect that the quagmire to which we are led by these issues is spawned by conflating the physical world with the mathematical formalism that is intended only to model it, but this is a topic for another conversation (Boughn 2019).

The purpose of this essay is neither to completely demystify quantum mechanics nor to stifle conversation about its interpretation. To be sure, the number of extraordinary quantum phenomena seems to be nearly without limit. Quantum spin, anti-matter, field theory, gauge symmetry, the standard model of elementary particles, etc., are all subsequent developments in quantum theory that have very little connection to classical physics and about which the above discussion has little to say. Certainly wave-particle duality is a mysterious fact of nature. Whether one considers it to be a fundamental principle, as did Bohr, or sees it as intimately related to the quantal character of the world is, perhaps, a matter of taste. I have sought to couch my discussion not in the mathematical formalism of quantum theory, but in terms of a simple physical principle: matter, radiation, and their interactions occur only in discreet quanta. Rather than quashing discussion about the meaning of quantum mechanics, perhaps this essay will stimulate new discussions.

**Acknowledgements**: I would like to thank Marcel Reginatto for many helpful conversations as well as for commenting on several early versions of this paper. Also, thanks to Serena Connolly for introducing me to Lucretius's wonderful poem.